\documentclass[preprint,aps,pra,showpacs,superscriptaddress]{revtex4}

\usepackage{graphicx}

\usepackage{amsmath}

\begin{document}

\preprint{}

\title{Some mathematical aspects in determining the 3D controlled solutions of the Gross-Pitaevskii equation}

\author{R. Fedele}
\email{renato.fedele@na.infn.it} \affiliation{Dipartimento di
Scienze Fisiche, Universit\`{a} Federico II and INFN Sezione di
Napoli, Complesso Universitario di M.S. Angelo, via Cintia,
I-80126 Napoli, Italy}
\author{D. Jovanovi\'c}
\email{djovanov@phy.bg.ac.yu} \affiliation{Institute of Physics,
P. O. Box 57, 11001 Belgrade, Serbia}

\author{S. De Nicola}
\email{s.denicola@cib.na.cnr.it} \affiliation{Istituto di
Cibernetica ``Eduardo Caianiello'' del CNR Comprensorio ``A.
Olivetti'' Fabbr. 70, Via Campi Flegrei, 34, I-80078 Pozzuoli
(NA), Italy} \affiliation{Dipartimento di Scienze Fisiche,
Universit\`{a} Federico II and INFN Sezione di Napoli, Complesso
Universitario di M.S. Angelo, via Cintia, I-80126 Napoli, Italy}

\author{B. Eliasson}
\email{bengt@tp4.rub.de} \affiliation{Institut f\"ur
Theoretische Physik IV, Ruhr--Universit\"at Bochum, D-44780
Bochum, Germany}\affiliation{Department of Physics,
Ume{\aa} University, SE-90 187 Ume{\aa}, Sweden}

\author{P. K. Shukla}
\email{ps@tp4.rub.de} \affiliation{Institut f\"ur Theoretische
Physik IV, Ruhr--Universit\"at Bochum, D-44780 Bochum, Germany}
\affiliation{SUPA, Department of Physics, University of Strathclyde, Glasgow
G4 ONG, United Kingdom}

\date{\today}

\begin{abstract}
The possibility of the decomposition of the three dimensional (3D)
Gross-Pitaevskii equation (GPE) into a pair of coupled
Schr\"{o}dinger-type equations, is investigated. It is shown
that, under suitable mathematical conditions, solutions of the 3D
controlled GPE can be constructed from the solutions of a 2D linear Schr\"{o}dinger equation
(transverse component of the GPE) coupled with a 1D nonlinear
Schr\"{o}dinger equation (longitudinal component of the GPE). Such a
decomposition, called the 'controlling potential method' (CPM), allows one
to cast the above solutions in the form of the product of the solutions of the
transverse and the longitudinal components of the GPE. The coupling
between these two equations is the functional of both the transverse
and the longitudinal profiles. The analysis shows that the
CPM is based on the variational principle that sets up a  condition on the 
controlling potential well, and whose physical interpretation is given
in terms of the minimization of the (energy) effects introduced by the
control operation.

\end{abstract}

\maketitle

\section{Introduction}\label{Introductory}
Since its experimental observation \cite{3}, the study of the three dimensional (3D) collective and
nonlinear dynamics of the Bose Einstein condensate (BEC) \cite{1} in
an external potential trap \cite{2} has received a great deal of attention by a very wide 
scientific community and in the investigations concerning fundamental physics, by mathematical physics
and sophisticated technological applications \cite{4}.
Although rapid scientific and technological
advances have been achieved in this area, finding the exact analytical 3D
solutions of the Gross-Pitaevskii equation (GPE) \cite{2}, that correspond to the
coherent state of a BEC in a suitable external potential well
(such as soliton-like structures), still remains a challenging task
for physicists and mathematicians.

A number of valuable approximative analytical \cite{5} and numerical evaluations \cite{6} have been 
presented in the literature and have been adequately compared with a very wide spectrum of experimental 
observations. The experience gained from these investigations may suggest the idea that a BEC's dynamics
exhibits the features of a nonlinear non-autonomous system
\cite{7} for which it seems to be necessary to include some control operations in order to allow the existence
of coherent structures. In particular, to
retain the 3D coherent stationary structures of the BEC for a long time, suitable "ad hoc" time-dependent 
external potentials and control operations are known to be necessary \cite{8}.
Furthermore, in the presence of
an inhomogeneous time-dependent external potential one encounters
some difficulties to find exact soliton solutions in one or more
dimensions, although several kind of solitons have been found in
certain approximations \cite{9}. Consequently, one easily arrives to the conclusion that, in order 
to get exact soliton structures, some sort of the 'control of the system' seems to be necessary. 
This implies that the correct analysis of the system should include a control potential term in 
the GPE which is to be determined dynamically by the system itself. In principle, this procedure may be
extended to an arbitrary 'controlled solution' with
the appropriate choice of the external potential (so-called
'controlling potential' \cite{10}\,). In fact, a
controlling potential method (CPM) has been recently proposed in the
literature and used to find multi-dimensional controlled localized
solutions of the GPE. In the preliminary investigations
\cite{11}, this method has established reasonable experimental control
operations that ensure the stability of the solution against relatively small
errors in the experimental realization of the prescribed controlling
potential. The main goal of the CPM is to fix the type of the  desired controlled solution and to find the 
appropriate family of the controlling potentials. Then, the set of suitable mathematical conditions has to 
be found allowing us to select the desired solution, with the adopted controlling potential.

In this manuscript, we develop an analytical procedure
to construct exact three dimensional solutions of a controlled
Gross-Pitaevskii equation, by improving the CPM. To this end, we
develop the theory of the BEC control based on two
\textit{decomposition theorems} leading to suitable physical
conditions to express the BEC wave function as the product of a 2D
wave function and a 1D wave function, taking into account the
'transverse' and 'longitudinal' BEC profiles, respectively.
Such a factorization allows us to decompose the 3D controlled GPE
into a set of coupled equations, comprising a 2D linear
Schr\"odinger equation (governing the evolution of the
'transverse' wave function), a 1D nonlinear Schr\"odinger equation
(governing the evolution of the 'longitudinal' wave function) and
a variational condition involving the controlling potential. The
requirement for the minimization of the effects introduced by the
control operations (i. e. the requirement that the average of the controlling potential
over the transverse plane is equal to zero) allows us to determine
explicitly the self-consistent controlling potential which also
plays the role of the coupling term between transverse and
longitudinal BEC dynamics.

\section{Controlled Gross-Pitaevskii equation}
\label{controlled-GPE} 
It is well known
that the spatio-temporal evolution of the ultracold system of
identical atoms forming a BEC in the presence of the external
potential $U_{ext}(\mathbf{r},t)$, within the mean field
approximation, is governed by the three dimensional
Gross-Pitaevskii equation \cite{2}, viz.,
\begin{equation}\label{GPE-0}
    i\mathop \hbar\frac{\partial \Psi }{\partial
t}=-\frac{\mathop \hbar }{2m_a}\nabla ^2\Psi + N Q\left| \Psi
\right|^2\Psi + U_{ext} \left(\mathbf{r},t\right)\,\Psi\,,
\end{equation}
where $\Psi (\mathbf{r},t)$ is the wavefunction describing the BEC
state, $m_a$ is the atom mass and $Q$ is a coupling coefficient
related to the short range scattering (s-wave) length $a$
representing the interactions between atomic particles, viz.,
$Q=4\pi\hbar^2a/m_a$, and N is the number of atoms. Note that the
short range scattering length can be either positive or negative.
We assume that $U_{ext}$ is the sum of the 3D trapping potential
well, $U_{trap}$, that is used to confine the particles of a BEC, and the
controlling potential $U_{contr}$ which will be determined
self-consistently. We conveniently
introduce the variable $s=ct$ ($c$ being the speed of light) and
divide both sides of Eq.(\ref{GPE-0}) by $m_a c^2$, and we use the notation
\begin{equation}\label{ext-potential}
  \frac{U_{ext}(\mathbf{r}, t)}{m_a c^2} = \frac{U_{trap}(\mathbf{r}, t)}{m_a c^2} +
  \frac{U_{contr}(\mathbf{r}, t)}{m_a c^2} \equiv V_{trap}(\mathbf{r},s) +
  V_{contr}(\mathbf{r},s)\,,
\end{equation}
Eq. (\ref{GPE-0}) can be cast in the form
\begin{equation}\label{3D-GPE}
    i\mathop \lambda \nolimits_c \frac{\partial \psi }{\partial
s}=-\frac{\mathop \lambda \nolimits_c^2 }{2}\nabla ^2\psi + \left[
V_{trap}(\mathbf{r},s) + V_{contr} \left(\mathbf{r},s\right) + q
|\psi|^2\right]\,\psi\,,
\end{equation}
where $\psi(\mathbf{r},s) \equiv \Psi (\mathbf{r},t=s/c)$,
$\mathop \lambda \nolimits_c \equiv \hbar/m_a c^2$ is the Compton
wavelength of the single atom of BEC and $q \equiv N Q/mc^2$.

In this paper, we will investigate the properties of Eq.
(\ref{3D-GPE}) and $V_{contr}$ that enable the existence of the controlled 3D
solutions in the factorized form
\begin{equation}\label{factorized-solution}
    \psi (\mathbf{r},s) = \psi_\bot(\mathbf{r}_\bot,s)\,\, \psi_z(\mathbf{r}_\bot,z,s)\,,
\end{equation}
provided that $V_{trap}$ can be split into two parts, as
\begin{equation}\label{V-trap}
    V_{trap}(\mathbf{r},s) = V_\bot (\mathbf{r}_\bot,s) + V_z (z,s)
\end{equation}
where, in Cartesian coordinates, $\mathbf{r} \equiv (x,y,z)$ and
$\mathbf{r}_\bot \equiv (x,y)$ denotes, by definition, the
'transverse' part of the particle's vector position $\mathbf{r}$.
We also refer to $z$ as to the 'longitudinal'
coordinate.

By substituting Eqs. (\ref{factorized-solution}) and
(\ref{V-trap}) in Eq. (\ref{3D-GPE}), we easily get:
{\small\begin{eqnarray}\label{factorized-3D-GPE}
 \psi_\bot \frac{\mathop \lambda \nolimits_c^2 }{2}\left[\nabla
_\bot ^2 \psi_z +2\,\frac{\nabla _\bot \psi _\bot}{\psi _\bot
}\cdot \nabla _\bot \psi_z\right] &+& \psi_\bot\,
  \left[i\mathop \lambda \nolimits_c \frac{\partial \psi _z }{\partial
s}+\frac{\mathop \lambda \nolimits_c^2 }{2}\frac{\partial
^2\psi_z}{\partial z^2} -\left(V_z + V_{contr}
+ q\left|\psi_\bot\right|^2 \left|\psi_z\right|^2\right)\psi_z\right]\nonumber\\
&+& \psi_z\,\left[i\mathop \lambda \nolimits_c \frac{\partial \psi_\bot }{\partial
s} + \frac{\mathop \lambda \nolimits_c^2
}{2}\nabla_\bot^2\psi_\bot - V_\bot (\mathbf{r}_\bot,s)
  \psi_\bot \right]  = 0\,,
\end{eqnarray}}
where, in Cartesian coordinates, $\nabla_\bot \equiv
\hat{x}\,\partial/\partial x + \hat{y}\,\partial/\partial y$.

Let us define as 'controlled parameter' the following
time-dependent quantity:
\begin{equation}\label{q1D-definition}
    q_{1D}(s) = q\,\int d^2\vec r_\bot\,
    \left|\psi_\bot\right|^4\,;
\end{equation}
and the following linear and nonlinear operators, respectively:
\begin{equation}\label{H-perp}
    \widehat{H}_\bot = - \frac{\mathop \lambda \nolimits_c^2 }{2}\nabla_\bot^2 + V_\bot (\mathbf{r}_\bot,s)
\end{equation}
\begin{equation}\label{H-z}
    \widehat{H}_z = - \frac{\mathop \lambda \nolimits_c^2 }{2}\frac{\partial^2}{\partial z^2} + V_z (z,s) +
    q_{1D}(s)\,|\psi_z (\mathbf{r}_\bot,z,s)|^2 + V_0
\end{equation}
where $V_0$ is an arbitrary real constant. Then, Eq.
(\ref{factorized-3D-GPE}) can be rewritten as:
\begin{eqnarray}\label{factorized-3D-GPE-1}
    \psi_z\,\left(i\mathop \lambda \nolimits_c \frac{\partial}{\partial
s} -\widehat{H}_\bot\right)\psi_\bot &+& \psi_\bot\,
  \left[\left(i\mathop \lambda \nolimits_c \frac{\partial}{\partial
s} -\widehat{H}_z \right)\psi _z +\left(q_{1D}(s) -
q|\psi_\bot|^2\right)|\psi_z|^2\psi_z
+ \left(V_0 - V_{contr}\right)\psi_z\right] \nonumber\\
&+& \psi_\bot \frac{\mathop \lambda \nolimits_c^2 }{2}\left[\nabla
_\bot ^2 \psi_z + 2\,\frac{\nabla _\bot \psi _\bot}{\psi _\bot
}\cdot \nabla _\bot \psi_z\right] = 0\,.
\end{eqnarray}

\section{The decomposition properties of the Controlled
Gross-Pitaevskii equation}\label{decomposition-ptoperties}
By the definition of the controlling potential, $V_{contr}$ depends both on $\psi_\bot$ and $\psi_z$. 
In particular, we assume
here that the space and time dependence of $V_{contr}$ is given
also through $\rho_\bot (\mathbf{r}_\bot,s) \equiv |\psi_\bot
(\mathbf{r}_\bot,s)|^2$, viz.,
\begin{equation}\label{V-contr-general}
    V_{contr} = V_{contr}\left(\rho_\bot (\mathbf{r}_\bot,s),z,s\right)\,.
\end{equation}
Moreover, defining also the following functional of $\rho_\bot$:
\begin{equation}\label{functional-V-cal}
    {\cal V}\left[\rho_\bot; z,s\right] = \int\,
    \rho_\bot (\mathbf{r}_\bot,s)\,V_{contr}
    \left(\rho_\bot
    (\mathbf{r}_\bot,s),z,s\right)\,d^2\mathbf{r}_\bot\,,
\end{equation}
the following theorem holds:

\medskip\medskip

\noindent \textbf{\small DECOMPOSITION THEOREM 1.}

\medskip

\noindent\textit{If}
\begin{equation}\label{1D-psi-z}
\psi_z (\mathbf{r}_\bot, z,s)= \psi_z (z,s),
\end{equation}
\textit{and} $\psi_\bot (\mathbf{r}_\bot,s)$ \textit{is the solution of the following 
2D linear Schr\"odinger equation}
\begin{equation}\label{2D-LSE}
   \left(i\mathop \lambda \nolimits_c \frac{\partial}{\partial
s} -\widehat{H}_\bot\right)\psi_\bot = 0\,,
\end{equation}
\textit{and} ${\cal V}$ \textit{is a stationary functional (with
respect to variations} $\delta \rho_\bot$ \textit{of}
$\rho_\bot$\textit{)}, \textit{assuming the value} ${\cal V}=
V_0$, \textit{conditioned by the constraints}
\begin{equation}\label{constraint-1}
    \int\,\rho_\bot\,d^2\mathbf{r}_\bot = 1\,,
\end{equation}
\textit{(normalization condition for $\psi_\bot$)}, \textit{and}
\begin{equation}\label{constraint-2}
    \int\,\rho_\bot^2\,d^2\mathbf{r}_\bot = \frac{q_{1D}(s)}{q} =
    \textit{given function}\,,
\end{equation}
\textit{then} $\psi_z$ \textit{is the solution of the following 1D
nonlinear Schr\"odinger equation}
\begin{equation}\label{1D-NLSE}
    \left(i\mathop \lambda \nolimits_c \frac{\partial}{\partial
s} -\widehat{H}_z \right)\psi _z = 0\,,
\end{equation}
\textit{and} $V_{contr}$ \textit{is given by}
\begin{equation}\label{V-contr-stationary}
    V_{contr}(\mathbf{r}_\bot,z,s) = \left[q_{1D}(s) - q
    |\psi_\bot (\mathbf{r}_\bot,s)|^2\right]|\psi_z (z,s)|^2 + V_0\,.
\end{equation}

\medskip
\noindent To prove this theorem, first of all, we note that the assumptions (\ref{1D-psi-z}) and
(\ref{2D-LSE}) allow us to reduce Eq. (\ref{factorized-3D-GPE-1}) to
\begin{equation}\label{reduced-eq}
\left(i\mathop \lambda \nolimits_c \frac{\partial}{\partial s}
-\widehat{H}_z \right)\psi _z +\left[q_{1D}(s) -
q|\psi_\bot|^2\right]|\psi_z|^2\psi_z + \left(V_0 -
V_{contr}\right)\psi_z =0\,.
\end{equation}
Secondly, the required stationarity of ${\cal V}$ with respect to
variations $\delta \rho_\bot$ of $\rho_\bot$ implies that
\begin{equation}\label{stationarity}
    \delta {\cal V} + \alpha
    (z,s)\,\delta\int\,\rho_\bot\,d^2\mathbf{r}_\bot+ \beta
    (z,s)\,\delta\int\,\rho_\bot^2\,d^2\mathbf{r}_\bot = 0\,,
\end{equation}
where $\alpha (z,s)$ and $\beta (z,s)$ are Lagrangian multipliers.
Taking into account Eq. (\ref{functional-V-cal}), condition
(\ref{stationarity}) allows us to solve the corresponding
ordinary inhomogeneous
first-order differential equation for $V_{contr}$ where
$\rho_\bot$ plays the role of the independent variable and $z$ and
$s$ are parameters, yielding the following general solution
\begin{equation}\label{V-contr-general-solution}
V_{contr}(\mathbf{r}_\bot,z,s) = \frac{h(z,s)}{\rho_\bot
(\mathbf{r}_\bot,s)} - \alpha (z,s) - \beta (z,s)\rho_\bot
(\mathbf{r}_\bot,s)\,,
\end{equation}
where $h(z,s)$ is an arbitrary function. Actually, to ensure the
convergence of the integral in the definition of the functional ${\cal V}$,
see Eq. (\ref{functional-V-cal}), it is easy to see that we must have $h(z,s) =
0$. Consequently, the appropriate $V_{contr}$ satisfying the
stationarity condition ${\cal V}=V_0$ is given by
\begin{equation}\label{V-contr-solution}
V_{contr}(\mathbf{r}_\bot,z,s) = \left[\frac{q_{1D}(s)}{q} -
\rho_\bot (\mathbf{r}_\bot,s)\right]\beta (z,s) + V_0\,,
\end{equation}
which after the substitution in Eq. (\ref{reduced-eq}) gives
\begin{equation}\label{reduced-equation-1}
    \left(i\mathop \lambda \nolimits_c \frac{\partial}{\partial s}
-\widehat{H}_z \right)\psi _z +\left[q_{1D}(s) -
q|\psi_\bot|^2\right]\left(|\psi_z|^2 - \beta/q\right)\psi_z =0\,.
\end{equation}
Now, according to the hypothesis (\ref{1D-psi-z}), to preserve the
$\mathbf{r}_\bot$-independence of $\psi_z$, Eq.
(\ref{reduced-equation-1}) can be satisfied only when
\begin{equation}\label{beta-solution}
\beta (z,s) = q|\psi_z (z,s)|^2\,,
\end{equation}
which immediately implies that Eqs. (\ref{1D-NLSE}) and
(\ref{V-contr-stationary}) are satisfied.

\medskip\medskip

\noindent \textbf{\small DECOMPOSITION THEOREM 2.}

\medskip

\noindent \textit{Let us suppose that} $\psi_z =\psi_z (z,s)$ \textit{is
the solution of the 1D nonlinear Schr\"odinger equation
(\ref{1D-NLSE}). Then, the functional} ${\cal V}$ \textit {given by
(\ref{functional-V-cal})\textit{and
conditioned by the constraints (\ref{constraint-1}), and
(\ref{constraint-2}), is stationary (with respect to
variations} $\delta \rho_\bot$ \textit{of} $\rho_\bot$\textit{),} ${\cal V}= V_0$  if, 
and only if,} $\psi_\bot = \psi_\bot
(\mathbf{r}_\bot,s)$ \textit{is the solution of the 2D linear
Schr\"odinger equation (\ref{2D-LSE})}.

\medskip

\noindent To prove this proposition, we observe that since $\psi_z
(z,s)$ satisfies Eq. (\ref{1D-NLSE}), Eq.
(\ref{factorized-3D-GPE-1}) becomes
\begin{equation}\label{reduced-factorized}
   \left(i\mathop \lambda \nolimits_c \frac{\partial}{\partial
s} -\widehat{H}_\bot\right)\psi_\bot +
  \left[\left(q_{1D}(s) -
q|\psi_\bot|^2\right)|\psi_z|^2 + \left(V_0 -
V_{contr}\right)\right]\psi_\bot = 0\,.
\end{equation}
By multiplying the latter on the left by $\psi_\bot^*$ and
integrating over all the transverse plane, we easily obtain
\begin{equation}\label{integrated-eq}
   \int\,\psi_\bot^*\left(i\mathop \lambda \nolimits_c \frac{\partial}{\partial
s} -\widehat{H}_\bot\right)\psi_\bot\,d^2\mathbf{r}_\bot +V_0 -
{\cal V}\left[\rho_\bot; z,s\right] = 0 \,,
\end{equation}
where constraints (\ref{constraint-1}) and (\ref{constraint-2})
have been used. Consequently, if $\psi_\bot$ satisfies Eq.
(\ref{2D-LSE}), then ${\cal V}$ is a stationary functional
with the value ${\cal V} = V_0$, conditioned by
(\ref{constraint-1}) and (\ref{constraint-2}). Conversely, the assumed stationarity of ${\cal V}$ 
implies that the functional
form of $V_{contr}$ with respect to $\mathbf{r}_\bot$, $z$ and $s$
is given by Eq. (\ref{V-contr-solution}), which substituted in Eq. (\ref{reduced-factorized}) gives
\begin{equation}\label{reduced-factorized-1}
   \left(i\mathop \lambda \nolimits_c \frac{\partial}{\partial
s} -\widehat{H}_\bot\right)\psi_\bot +
  \left(q_{1D}(s) - q|\psi_\bot|^2\right)\left(|\psi_z|^2 - \beta (z,s)/q\right) = 0\,.
\end{equation}
However, if $\beta (z,s)/q \neq |\psi_z (z,s)|^2$, then $\psi_z$ would
be also function of $\mathbf{r}_\bot$ which would
contradict the assumption $\psi_z = \psi_z (z,s)$. It follows that
$\beta (z,s)/q = |\psi_z (z,s)|^2$ and, in turn, that Eq.
(\ref{2D-LSE}) is satisfied.

The results presented above allow us to draw the following
conclusion.

\medskip

\noindent\textit{If} $\psi_\bot (\mathbf{r}_\bot,s)$ \textit{and}
$\psi_z (z,s)$, \textit{are two complex functions which are exact
solutions of the 2D linear Schr\"odinger equation (\ref{2D-LSE})
and the 1D nonlinear Schr\"odinger equation (\ref{1D-NLSE}),
respectively, provided that }$V_{contr}$ \textit{is given by Eq.
(\ref{V-contr-stationary}), the function $\psi (\mathbf{r},s) =
\psi_\bot (\mathbf{r}_\bot,s)\,\psi_z (z,s)$ is the exact solution
of the controlled 3D Gross-Pitaevskii equation (\ref{3D-GPE})}.

Of course, the inverse is not necessarily true. In fact, it is
easy to see that, in principle, it is not true that an arbitrary
solution of Eq. (\ref{3D-GPE}) can be expressed as the product of
two wave functions $\psi_\bot (\mathbf{r}_\bot,s)$ and $\psi_z
(z,s)$ that obey the Eqs. (\ref{2D-LSE}) and (\ref{1D-NLSE}),
respectively. In other words, we can decompose the controlled 3D
GPE (\ref{3D-GPE}) into the system of equations (\ref{2D-LSE}),
(\ref{1D-NLSE}) and (\ref{V-contr-stationary}) only for the subset of its solutions 
of the type (\ref{factorized-solution}). However, using
such a decomposition we are able to solve Eq.
(\ref{3D-GPE}) and to obtain a wide spectrum of exact solutions of the type
(\ref{factorized-solution}).

\section{Conclusions and Remarks}

In this paper, we have presented some mathematical properties of the controlled 3D GPE (\ref{3D-GPE}). 
After formulating and proving two decomposition theorems, we have found the mathematical conditions 
that make possible the construction of the solution in a factorized form, i.e. $\psi (\mathbf{r},s) 
= \psi_\bot(\mathbf{r}_\bot,s)\,\, \psi_z(z,s)$, where $\psi_\bot(\mathbf{r}_\bot,s)$ and
$\psi_z(z,s)$ satisfy the 2D linear Schr\"{o}dinger equation
($i\mathop \lambda \nolimits_c {\partial \psi_\bot}/{\partial
s} =\widehat{H}_\bot\psi_\bot $) and the nonlinear controlled nonlinear Schr\"{o}dinger 
equation ($i\mathop \lambda \nolimits_c {\partial\psi _z}/{\partial
s} =\widehat{H}_z\psi _z $), respectively.
The results presented here improve the formulation of the recently proposed Controlling Potential Method \cite{10,11}.

It is worthy observing that the set of equations (\ref{2D-LSE}),
(\ref{1D-NLSE}) and (\ref{V-contr-stationary}) opens up the
possibility to find the controlled solutions of the type
(\ref{factorized-solution}) which exhibit the quantum character in
the transverse part (superposition principle with consequent
interference effects) and the classical character in the
longitudinal part (due to the nonlinearity of the 1D nonlinear
Schr\"odinger equation), although the entire solution of the
controlled 3D GPE is nonlinear and, therefore, has a
classical character. By means of suitable controlling
and trapping potentials, this possibility would allow, for
instance, for a very stable soliton-like longitudinal profile of
the BEC whose transverse profile would have a quantum
character as a result of the quantum interference at the macroscopic level.

Note that, when $\psi_\bot $ satisfies Eq. (\ref{2D-LSE}),
according to definition (\ref{functional-V-cal}), ${\cal V}$
represents the average of $V_{contr}$ in the transverse plane. The
value of this average corresponds to the arbitrary constant $V_0$.
Without loss of generality, we put $V_0 = 0$, viz.
\begin{equation}\label{mimimize_2-D}
\int d^2\vec r_\bot \,\, \psi_\bot^*\, V_{contr}\,\psi_\bot =0 .
\end{equation}
This way, among all possible choices of $V_{contr}$, we adopt the
one which does not change the mean energy of the system
(note that the
average of the Hamiltonian operator in Eq. (\ref{3D-GPE}) is the
same with or without $V_{contr}$) and thus minimizes the
effects introduced by our control operation.

In our forthcoming papers, we will use the method developed in the present paper to solve
exactly the 3D controlled GPE with a 3D parabolic potential trap. We find the controlled envelope 
solutions in the form of localized as
well as periodic structures for which suitable stability analysis
is performed.

\acknowledgments
This work was partially supported by Fondo Affari Internazionali of Istituto Nazionale di Fisica Nucleare, Sezione di Napoli, (Napoli, Italy), by the Deutsche Forschungsgemeinschaft
(Bonn, Germany) through the project SH21/3-1 of the Research Unit 1048, and by the 
Swedish Research Council (VR).

\end{document}